\documentclass[conference]{IEEEtran}
\IEEEoverridecommandlockouts
\usepackage{subcaption}  
\usepackage{color}
\usepackage{pifont}
\usepackage{bbm}
\usepackage{stfloats}
\usepackage[short]{optidef}
\usepackage{multirow}
\usepackage{comment}
\usepackage{cite}
\usepackage{amsmath,amssymb,amsfonts}
\usepackage{algorithmic}
\usepackage{graphicx}
\usepackage{textcomp}
\usepackage{xcolor}

\def\BibTeX{{\rm B\kern-.05em{\sc i\kern-.025em b}\kern-.08em
    T\kern-.1667em\lower.7ex\hbox{E}\kern-.125emX}}
\begin{document}

\title{
A Two-Stage CAE-Based Federated Learning Framework for Efficient Jamming Detection in 5G Networks
}

\author{\IEEEauthorblockN{Samhita Kuili, Mohammadreza Amini, Burak Kantarci} 
\IEEEauthorblockA{\textit{School of Electrical and Computer Engineering} \\
\textit{University of Ottawa}\\
Ottawa, Canada \\
Emails: \{skuil016, mamini6, burak.kantarci\}@uottawa.ca}

\vspace{-0.2in}}

\maketitle

\begin{abstract}
Cyber-security for 5G networks is drawing notable attention due to an increase in complex jamming attacks that could target the critical 5G Radio Frequency (RF) domain. These attacks pose a significant risk to heterogeneous network (HetNet) architectures, leading to degradation in network performance. Conventional machine-learning techniques for jamming detection rely on centralized training while increasing the odds of data privacy. To address these challenges, this paper proposes a decentralized two-stage federated learning (FL) framework for jamming detection in 5G femtocells. Our proposed distributed framework encompasses using the Federated Averaging (FedAVG) algorithm to train a Convolutional Autoencoder (CAE) for unsupervised learning. In the second stage, we use a fully connected network (FCN) built on the pre-trained CAE encoder that is trained using Federated Proximal (FedProx) algorithm to perform supervised classification. Our experimental results depict that our proposed framework (FedAVG and FedProx) accomplishes efficient training and prediction across non-IID client datasets without compromising data privacy. Specifically, our framework achieves a precision of 0.94, recall of 0.90, F1-score of 0.92, and an accuracy of 0.92, while minimizing communication rounds to 30 and achieving robust convergence in detecting jammed signals with an optimal client count of 6.

\end{abstract}

\begin{IEEEkeywords}
5G, Federated Learning, Jamming Detection, Convolutional Autoencoder, Non-IID data, Over-The-Air Transmission

\end{IEEEkeywords}

\section{Introduction}

The rapid growth in usage of intelligent user devices demands enhanced spectrum efficiency (SE) and fast data transmission in fifth-generation (5G) networks. In this direction, wireless networks have advanced into heterogeneous networks (HetNets) to provide reliable services to multiple end-users. HetNets provision dense deployment of small cells to collaborate effectively in macrocell, which augments the SE and system throughput of the wireless network. These small cells are commonly known as femto cells. Each femto cell comprises small base station which transmit low power to improve the quality of service (QoS) requirements for the services availed by user devices in a 5G wireless network. The ability of broadcasting channel spectrum of 5G wireless networks is susceptible to security attacks: jamming which causes performance degradation in the network \cite{pirayesh2022jamming}. Hence, different approaches have been proposed to detect and mitigate such attacks \cite{asemian2025active}. The 5G network ensures high security and robustness to jamming attacks compared to Long-Term Evolution (LTE) networks or 4G networks \cite{arjoune2020smart}. Additionally, each layer of 5G-NR protocol stack comprises attack surfaces, increasing the odds of vulnerability to jamming attacks. Therefore, causing a bottleneck in communication overhead. Various strategies have been implemented to detect jamming attacks, broadly categorized into non-machine and machine learning (ML)-based approaches. Among ML-based methods, Federated Learning (FL) stands out as a promising technique, as it aims to develop a robust global model by integrating diverse observations from differently-configured femtocells while preserving user data privacy. This paper presents an efficient way of creating an FL-based jamming detector in 5G networks using domain knowledge information. 
In particular, we propose a two-stage CAE-based federated learning framework for jamming detection in a heterogeneous environment, exploiting both unsupervised and supervised learning. The proposed structure leverages the strengths of unsupervised jamming detection combined with supervised fine-tuning, while ensuring data privacy is preserved. Unlike many studies, the global model is trained over different real-world data sets collected from the 5G TELUS network. Leveraging 5G domain knowledge, we use a crucial part of the 5G resource grid, namely the Synchronization Signal Block (SSB). This involves processing over-the-air I-Q samples and extracting 4 OFDM symbols related to SSB\footnote{{Synchronization in the time-frequency domain is a crucial process enabling 5GNR user equipment
(UE) to effectively send and receive data. A jamming attack
in this step can effectively disrupt the communication link.}}.
The main contributions of the paper are summarized below:

\begin{enumerate}
    \item Develop a decentralized two-stage CAE-based federated learning system for jamming detection in the 5G RF domain. The strategy includes two phases: an unsupervised learning process based on signal reconstruction, and a supervised learning process through a classification layer. 
    \item Achieve a high-performance global model by selecting an optimal set of clients. The client selection process not only helps reduce communication rounds but also ensures the global model remains unbiased.
\end{enumerate}


\vspace{-0.25cm}
\section{Related Works}\label{RW}

Conventional ML algorithms are exploited to detect jamming attacks, yet they rely on a centralized model training, which increases network load 
and a greater risk of data leakage while  training 
on spectrum channel shared between user/client devices and femto base station.  On the contrary, FL is a decentralized paradigm that trains a model across multiple clients using data parallelism \cite{dean2012large}, while each client retains data locally, addressing critical factors of computational capacity, privacy, and security issues to the data \cite{zhu2021federated}. However, FL performance often degrades with non-independent and identically distributed (non-IID) data, a challenge prevalent in real-world 5G-NR wireless networks \cite{mcmahan2017communication}. Jamming attacks are typically malicious attacks launched by an adversary to cause intentional interference in 5G wireless cellular network \cite{Lohan.2024}. These attacks in 5G NR can be categorized into constant jammer, deceptive jammer, random jammer, reactive jammer, Go-next jammers, and control channel jammers\cite{arjoune2020smart, wesolowski2023simple}. Mao et al. \cite{mao2023deep} highlight a thorough review on the usage of deep learning on PHY layer in the context of 5G and 6G networks but lacks coverage of security aspects. Varotto et al. \cite{varotto2024detecting} adopt CNN, and other machine learning techniques to detect jamming attack or a SSB jammer on the narrowband of 5G network, 
demonstrating a comparative classification performance of jammed and non-jammed signals.
FL is broadly classified into two categories: horizontal federated learning 
and vertical federated learning. In this work of jamming detection, we prioritize the horizontal FL approach where the training dataset for each local client indicates fixed feature space and shares different sample sizes or observations. 
Zhu et al. \cite{zhu2021federated} provide  
a detailed survey on the implication of non-IID data on parametric models which might result in global model divergence based on the distribution of local datasets while training FL model framework. 
Considering the above issue that persists in FL, selecting the optimal number of clients 
forms an important course of action while analyzing the performance of FL.  
Gouissem et al. \cite{gouissem2023comprehensive} provide a meticulous review of existing state-of-the-art approaches to client selections in federated learning. Moreover, random selection of clients is a conventional method often followed in FL. McMahan et al. \cite{mcmahan2017communication} highlight the efficiency of the FedAvg algorithm across various deep neural network architectures, showing that it performs well with fewer communication rounds compared to the FedSGD baseline. Their approach assumes a fraction of clients per round to boost computational parallelism and increase local computations per client. 
Another approach is to adopt a performance-based selection of clients. Ribero and Vikal in \cite{ribero2020communication} propose 
 fixed threshold and adaptive threshold strategies for client selection, 
 reducing communication rates by selecting only specific client weights 
 modeled as weight vectors using Ornstein-Uhlenbeck (OU) stochastic processes during Stochastic Gradient Descent (SGD). 
Sahu et al. \cite{Sahu2018FederatedOI} introduced FedProx, a framework that generalizes and re-parameterizes FedAvg to improve convergence stability when working with non-IID datasets. Zheng et al. \cite{zheng2023federated} leverage FedProx algorithm and introduce a joint client selection with unreliable communication and heterogeneous networks, which aims to accelerate the convergence. 

\begin{figure*}[ht]
\centerline{\includegraphics[scale = 0.277]{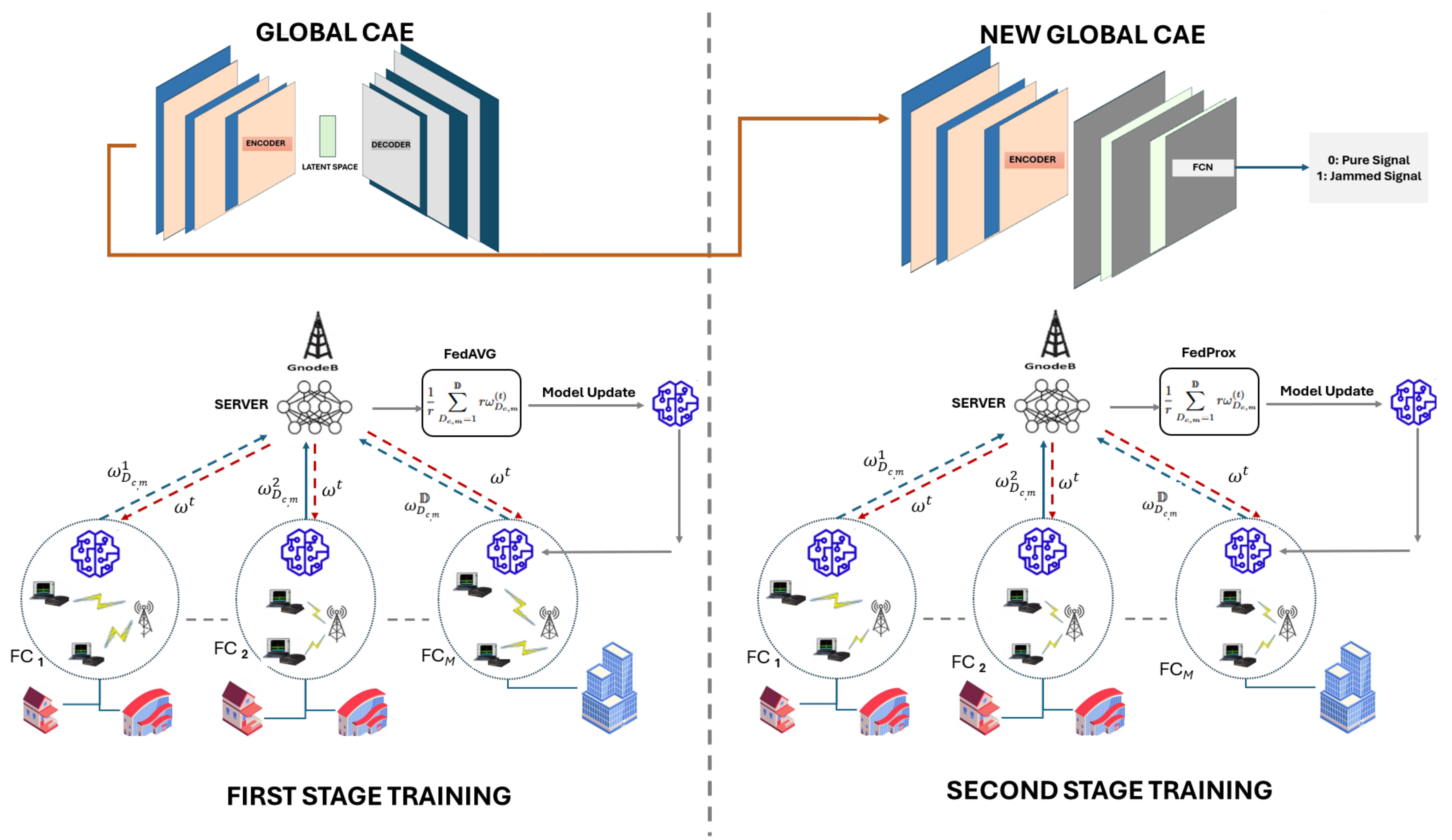}}
\caption{Jamming detection in a 5G-NR cellular network using two-stage federated leaning across multiple femtocells}
\label{fig1} \vspace{-4mm}
\end{figure*}


\section{SYSTEM MODEL}\label{M}
A FL scenario in a 5G network consisting of $M$ femtocells each serving multiple clients is considered. A jammer exists in the network, attempting to disrupt the communication links. To detect the jamming attack, the clients use SSB. In 5G NR systems, each radio cell is distinguished by a cell ID from a pool of 1008 IDs, organized into 336 distinct groups. Each group is designated by the cell ID group, $N_{ID}^1 \in \{0,1,\cdots,355\}$. It also consists of three different sectors specified by the cell ID sector $N_{ID}^2 \in \{0,1,2\}$. These IDs can be detected by UE from Secondary Synchronization Signal (SSS) and Primary Synchronization Signal (PSS) respectively. Then, the serving cell ID, (i.e. Physical Cell ID (PCI) is calculated as $N_{ID}^{cell}=3*N_{ID}^1+N_{ID}^2$.
Let $s(n)$ be the $n^{th}$ I-Q sample of the SSB signal transmitted by gNodeB (gNB) which can be represented as \vspace{-4mm}

\begin{equation}\label{Eq_tr_wave}
\begin{split}
    s(j)=\sum_{l=0}^{3} s_{l}(m) \quad     j=0,1, \cdots, (l \times m-1) \, 
\end{split}
\end{equation}
\noindent where $s_{l}(m)$, $m \in \{0,1, \cdots, N_{F\hspace{-.6mm}F\hspace{-.6mm}T} -1\}$, is the $m^{th}$ data symbol of $l^{th}$ SSB OFDM symbol and $N_{F\hspace{-.6mm}F\hspace{-.6mm}T}$ is the size of FFT. Each OFDM symbol $s_l(m)$ contains some data symbols $S_{l,k}$ in the frequency domain which is transformed into time domain as,  \vspace{-4mm}

\begin{equation}\label{Eq_ifft}
\begin{split}
    s_{l}(m)=\frac{1}{N_{F\hspace{-.6mm}F\hspace{-.6mm}T}} \sum_{k=0}^{N_{F\hspace{-.6mm}F\hspace{-.6mm}T}-1} S_{l,k} \,e^{{j2\pi k m}/{N_{F\hspace{-.6mm}F\hspace{-.6mm}T}}} \, 
\end{split}
\end{equation}

The PSS, which is the first OFDM symbol of SSB, i.e. $s_{l}(m) \mid_{l=0}$,  comprises one of three 127-symbol m-sequences and is assigned to the first symbol of each SSB, covering 127 subcarriers. The three potential m-sequences for the PSS are defined as follows \cite{Omri2019}.  \vspace{-4mm}

\begin{equation} \label{Eq_PSS}
    \begin{split}
       S_{l,k+i}\mid_{l=0}=\begin{cases}
       1-2d_p(i) \quad  k \in \{56, \cdots, 182 \}\\
       0      \quad \quad \quad \quad  Otherwise  , 
       \end{cases}
    \end{split}
\end{equation}
\noindent where $d_p(i)$ represents the m-sequences which are given in the 3GPP standard \cite{3gpp.38.211}.

Similar to LTE, 5G NR SSS serves to detect the physical cell identity. In contrast, the SSS comprises one of 336 127-symbol gold sequences, specifically assigned to the third symbol of each SSB. The 336 potential gold sequences for the SSS are outlined as follows.  \vspace{-4mm}

\begin{equation} \label{Eq_PSS_1}
    \begin{split}
       X_{l,k+i}\mid_{l=3}&=       \big[1-2d_s(i+k_0) mod \, 127\big] \\ 
       & \hspace{5mm}\times \big[ 1-2 d^{\prime }_s (i+k_1) \, mod \, 127  \big] \\
       & \hspace{5mm}  k \in \{56, \cdots, 182 \} \,  ,
    \end{split}
\end{equation}
\noindent where $k_0$ and $k_1$ are derived as,  \vspace{-4mm}

\begin{equation} \label{Eq_PSS2}
    \begin{split}
       k_0=15\Bigg[\frac{N_{ID}^1}{112} \Bigg] +5 N_{ID}^2 \,  ,\\
       k_1=N_{ID}^1 \, mod \, 112 \,  .
    \end{split}
\end{equation}

Furthermore, $d_s(i)$ and $d^{\prime}_s(i)$ can be extracted recursively as stated in 5G NR standard  \cite{3gpp.38.211}.

At the receiver, the perfect (without jamming) received SSB signal is expressed as,  \vspace{-4mm}

\begin{equation}\label{Eq_rec_wav}
\begin{split}
    x(j)= s(j) \circledast h(j) + w(j)   \,  ,  
\end{split}
\end{equation}
where $h(j)$ is the channel impulse response and $w(j)$ is the environmental noise. In the presence of a jammer, the received SSB signal is expressed as,  \vspace{-4mm}

\begin{equation}\label{Eq_rec_wav}
\begin{split}
    x(j)= s(j) \circledast h(j) + w(j) + s_J(j)   \,  ,  
\end{split}
\end{equation} where $s_J(j)$ is the jamming signal.

Similar to other detection problems, jamming detection can be formulated as a binary hypothesis test with $H_0$ and $H_1$ as null and alternate hypotheses respectively. In this context, the null hypothesis signifies the absence of the jammer. Let $x_{c,m}^n$ be the $\textit{n}^{th}$ observation (the $\textit{n}^{th}$ received SSB) in the $\textit{c}^{th}$ client data set in the $\textit{m}^{th}$ femtocell\footnote{For the sake of simplicity, index $j$ is dropped}.

The FL framework shown in Fig. \ref{fig1} is assumed which consists of multiple femtocells with 5G femto base stations surrounded by clients at varying geographical locations. Consequently, our framework consists of one server and $\mathbb{D}$ clients across all femto cells. 
Each femtocell \( F_m \) (\( m = 1, 2, \dots, M \)) contains a distinct set of $\mathcal{D}_m$ client datasets. 
Each $\mathcal{D}_m$ inherits a set of local datasets of RF domain and is denoted as
$\mathcal{D}_m = \{D_{1,m}, D_{2,m}, \dots, D_{C_m,m}\}$, 
where $D_{c,m}$ ($c \in \{1, \cdots, C_m\}$) represent $c^{th}$ client datasets in $m^{th}$ femtocell and $C_m$ is the total number of clients in that femtocell. 
The total number of client datasets across all femtocells can be represented as \vspace{-3mm}

\begin{equation}
    \mathbb{D} = \bigcup^{M}_{m=1} \mathcal{D}_m
\end{equation}
In addition, the size of the dataset ${D_{c,m}}$ can be represented as ${D_{c,m}}$ = \{($x_{c,m}^1$, $y_{c,m}^1$), ($x_{c,m}^2$, $y_{c,m}^2$),$\cdots$,($x_{c,m}^r$, $y_{c,m}^s$)\}, where $r$ $\in$ $\mathbb{R}^{P \ast Q}$ denotes the size of training data with \textit{P} SSB observations and \textit{Q} as length of IQ samples and $s$ $\in$ $\mathbb{R}^{P \ast 1}$ is the size of the ground truth of each client dataset.

\subsection{Two-stage federated learning on jamming detection}
In federated learning, weights and parameters obtained for the model play a vital role in estimating the performnace of the model in detection of jammed signals across the number of clients. In addition, a two-stage federated learning process involves deploying a reconstruction and classification module. The Convolutional Autoencoder (CAE) is employed in the first stage of the process, which is trained over training samples of SSB and IQ observations without any ground truth. This causes the extraction of the weights associated with encoder, latent space, and decoder. However, the trained weights of the encoder are further exploited in the second stage of federated learning, which assists a fully connected network in performing a classification module. 

In the first stage of FL, the server broadcasts the global model CAE to local clients for each communication round. Each client dataset  ${D_{c,m}}$ trains the model in an unsupervised manner to compute the reconstructed weights or local weights $\omega_{{D_{c,m}}}$ learned over each iteration in FL. After local training, the clients transmit the local weights to the server. The server aggregates the local weights and forms an average weight, $\omega$. Furthermore, the server broadcasts the updated global weights to be leveraged by the clients for the next round of iteration in FL. The reconstruction mean squared error loss (MSE) 
is denoted as $f(\psi, x_{c,m}^r) = \frac{1}{2} \| x_{c,m}^r - f_{nn}(x_{c,m}^r; \psi) \|^2$, where $f_{nn}(x_{c,m}^r; \psi)$ is the reconstructed output of input $x_{c,m}^r$ in CAE and $\psi$ is the weight vector to be learned. Let us assume the loss function for the client dataset ${D_{c,m}}$, which estimate the model error on its dataset as 
\vspace{-4mm}

\begin{equation}\label{eqn1}
    F_{{D_{c,m}}}(\omega) = \frac{1}{r} \sum_{e=1}^{r} \sum_{f=1}^{Q} \left( x_{{{D_{c,m}}},e,f} - f_{nn}(x_{{{D_{c,m}}},e,f}; \omega) \right)^2
\end{equation}

With the adoption of stochastic gradiet descent (SGD), the local weight of client ${D_{c,m}}$ at time $t$ is computed as \vspace{-2mm}

\begin{equation}\label{2}
    \omega_{{D_{c,m}}}^{(t)} = \omega^{(t-1)} - \eta \nabla F_{{D_{c,m}}} \left( \omega^{(t-1)} \right)
\end{equation}

where $\eta$ is the learning rate to train the local model for each client. In a typical FL scenario, each client would compute its own local weight through local training of the CAE, and further transmit the updated weight to the server. Moreover, the server will perform aggregation of all the received local weights to compute the global update using \vspace{-4mm}

\begin{equation}\label{3}
    \omega^{(t)} = \frac{1}{r} \sum_{{D_{c,m}}=1}^{\mathbb{D}} r \omega_{{D_{c,m}}}^{(t)}.
\end{equation}

Moreover, the global update is broadcasted back to each client and finally compute the global update by combining (\ref{2}) and (\ref{3}) into (\ref{4}) \vspace{-4mm}

\begin{equation}\label{4}
    \omega^{(t)} = \omega^{(t-1)}- \frac{\eta}{r} \sum_{{D_{c,m}}=1}^{\mathbb{D}}  r\nabla F_{{D_{c,m}}} \left( \omega^{(t-1)} \right).
\end{equation}

Consequently, the computation of global MSE loss comprising all $\mathbb{D}$ clients is represented as  \vspace{-4mm}

\begin{equation}\label{5}
F_{\text{global}}(\omega) = \frac{1}{\sum_{{D_{c,m}}=1}^{\mathbb{D}} r_{{D_{c,m}}}} \sum_{{D_{c,m}}=1}^{\mathbb{D}} r_{{D_{c,m}}} \cdot F_{{D_{c,m}}}(\omega)
\end{equation}

According to the proposed architecture for CAE, the weights obtained are further exploited in the second stage of evaluation. In first stage of FL, $\psi$ represents the weights vector learned over the communication rounds through the help of optimizer stochastic gradient descent while minimizing the loss function and achieving convergence. Therefore, a global CAE model $\mathcal{M}_{ENC+DEC}$ is trained at the server by obtaining the updated local CAE model trained across multiple clients. 
However, the second stage of FL instantiates a binary classification problem by adding an FCN followed by a few dense layers as a classifier head to the encoder of CAE. We intend to bring the pre-trained encoder $\mathcal{E}_{pre}$ from the global model CAE and use it as a feature extractor by freezing its weights $\varphi$. A fully connected layer (FCN) is adopted with $\mathcal{E}_{pre}$ to attain a new global CAE  model $\mathcal{M}_{ENC+FCN}$ acting as a classifier, and enable $\mathcal{M}_{ENC+FCN}$ to train over ${D_{c,m}}$. The global model $\mathcal{M}_{ENC+FCN}$ is further broadcasted to local clients and receives the updated weights from the models trained locally after each round of communication which also ensures tracking of the detection ability of jammed signals aiming to acquire the best optimal clients. 

During the training process of the first stage of proposed framework, the global base station (server) leverages the FedAVG \cite{mcmahan2016federated} algorithm as aggregation method, which aggregates the model weights from the local models and further updates the global model at the server to establish a new global model $\mathcal{M}_{ENC+DEC}$. In the end, the server shares the updated global model with the local clients. Moreover, the training is unsupervised capturing the fine-grained IQ samples of each SSB observation. This encourages the $\mathcal{M}_{ENC+DEC}$ to comprehend the underlying knowledge of each local client 
\begin{figure}[t]
\centerline{\includegraphics[scale = 0.52]{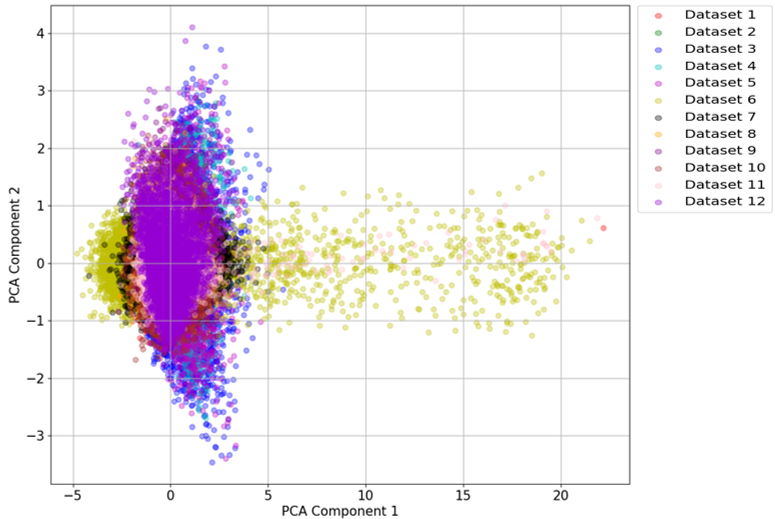}}
\caption{PCA distribution of Non-I.I.D Datasets}
\label{fig2} \vspace{-4mm}
\end{figure}
showcasing variation and deviation in attribute values corresponding to all SSBs and the balanced class information of jammed and pure signals existing across all client datasets. On the contrary, training process in the second stage of FL involves incorporating regularization and re-parametrimization of FedAVG; by adopting FedProx \cite{li2020federated} aggregation method. FedProx achieves convergence while dealing with heterogenous non-IID distribution datasets. On each iterative round of training, more clients will start participating locally to update the global model. This causes exposure to statistical non-IID distribution, leading to divergence in the local updates with additional clients involved in each round of training. FedProx introduces a proximal term that limits the deviation of local updates and restricts it closer to the global model between communication rounds. Specifically, instead of updating the model weights via minimizing the objective function $F_{{D_{c,m}}}(\omega)$ in (\ref{eqn1}), client 

\begin{equation}
    \min_{w} g_l(w; w^t) = F_{{D_{c,m}}}(\omega) + \frac{\mu}{2} \|w - w^t\|^2
\end{equation}

FedAVG is a special case of FedProx with  $\mu = 0$ and with the local solver chosen to be SGD. For the implementation of FedProx, we intend to leverage Binary cross-entropy loss function and the usage of optimizer adaptive moment estimation (Adam).

\section{Experiments}\label{E}

To address the issue of jamming detection in 5G RF domain using FL approach, we perform collection of real-world IQ samples over-the-air 5G transmission at multiple locations. In addition, we propose a two-stage federated learning approach to unravel the intrinsic details underlying across all heterogeneous datasets. Moreover, prior to execution of two stage federated learning approach, we implement a search for the presence of non-IID statistical distribution across all datasets by leveraging principal component analysis (PCA) (Fig. \ref{fig2}). 

\begin{table}[t]\vspace{0.17cm}
\caption{Parameters/Hyperparameters for the first stage: FedAVG}
\label{table:fed_params}\vspace{-0.145cm}
\centering
\resizebox{5.5cm}{!}{
\begin{tabular}{|c|c|}
\hline
\textbf{Parameter/ Hyperparameter} & \textbf{Value/Setting} \\ \hline
\textbf{Number of Clients}        & \{12,6\}                    \\

\textbf{Batch Size}               & 64                    \\ 
\textbf{Number of Rounds}         & 15                    \\ 
\textbf{Model}                    & CAE            \\ 
\textbf{Optimizer}         & SGD                   \\ 
\textbf{Learning Rate}     & 0.001                 \\ 
\textbf{Loss Function}            & Mean Squared Error (MSE) \\ 
\textbf{Training Data}            & X\_train, Y\_train    \\ 
\textbf{Validation Data}          & X\_valid, Y\_valid    \\ \hline

\end{tabular}
}\vspace{-0.3cm}\
\end{table}

\begin{table}[t]
\caption{Parameters/Hyperparameters of the second stage: FedProx}
\label{table:fed_params_}\vspace{-0.145cm}
\centering
\resizebox{5.5cm}{!}{
\begin{tabular}{|c|c|}
\hline
\textbf{Parameter/Hyperparameter} & \textbf{Value/Setting} \\ \hline
\textbf{Number of Clients}        & \{12,6\}                    \\

\textbf{Batch Size}               & 200                    \\ 
\textbf{Number of Rounds}         & 30                   \\ 
\textbf{Model}                    & CAE            \\ 
\textbf{Optimizer}         & Adam                   \\ 
\textbf{Learning Rate}     & 0.001                 \\ 
\textbf{Proximal term}         & 0.01                   \\ 
\textbf{Loss Function}            & Binary Cross Entropy (BCE) \\ 
\textbf{Training Data}            & X\_train, Y\_train with labels    \\ 
\textbf{Validation Data}          & X\_valid, Y\_valid with labels     \\
\textbf{Testing Data}          & X\_test, Y\_test with labels
\\ \hline

\end{tabular}
}\vspace{-0.3cm}\
\end{table}

\textbf{Dataset and Model}: The real world RF domain dataset consists of SSB observations each with sufficient IQ samples. In order to avoid misclassification, the datasets used for training and evaluation tasks reflect a balanced class of pure (0) and jammed (1) signals. Each dataset contains 5000 SSB  and 3297 IQ samples as training samples; with 2500 as class 0 and 2500 as class 1.
 We assume a train set of 3600 samples, a validation set of 400 samples, and a test set of 1000 samples by considering 70:10:20 split ratio for each dataset. Additionally, we adopt a Convolutional Autoencoder (CAE), which contains encoder of three layers with number of neurons [512, 256, 128] and decoder with number of neurons [128, 256, 512] with a dropout of 0.2 and a ReLU activation function.  During the first stage of FL, the parameters and hyperparameters chosen for undergoing the unsupervised learning on the samples based on the split ratio are highlighted in Table \ref{table:fed_params}. Furthermore, the initiation of second stage involves using the pre-trained encoder $\mathcal{E}_{pre}$ and the new global CAE model $\mathcal{M}_{ENC+FCN}$ which includes similar dimension of layers and neurons for encoder with additional dimension of dense layers of neurons [1024, 512, 256, 1] for classifier, an activation ReLU and a dropout of 0.5 for each dense layers and Sigmoid at the output layer, setting the $\mathcal{E}_{pre}$ layers frozen and enabling only the classifier trainable. Similarly, parameters and hyperparameters selected for the second stage of FL are given in Table \ref{table:fed_params_}. We run our proposed method on 12 clients and 6 clients while maintaining similar parameters and hyperparameters. In addition, we provide a comparison of proposed method over baseline algorithms, FedAVG and FedProx, applied to all and a subset of clients.

\section{EXPERIMENTAL RESULTS}\label{ER}
\begin{figure}[t]
\centerline{\includegraphics[scale = 0.21]{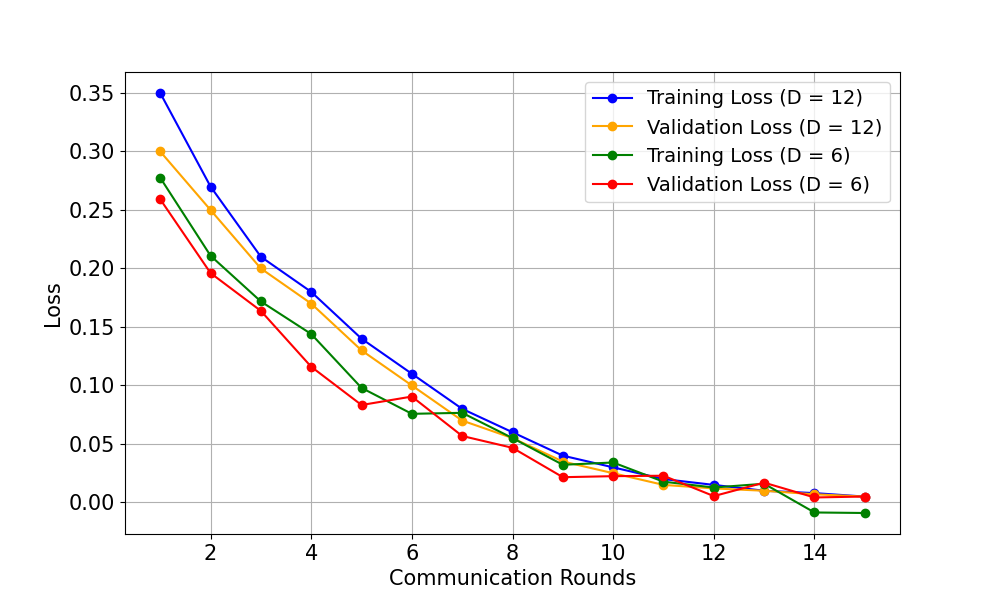}}
\caption{MSE loss convergence over communication rounds}
\label{fig3}\vspace{-2mm}
\end{figure}

\begin{figure}[t]
\centerline{\includegraphics[scale = 0.21]{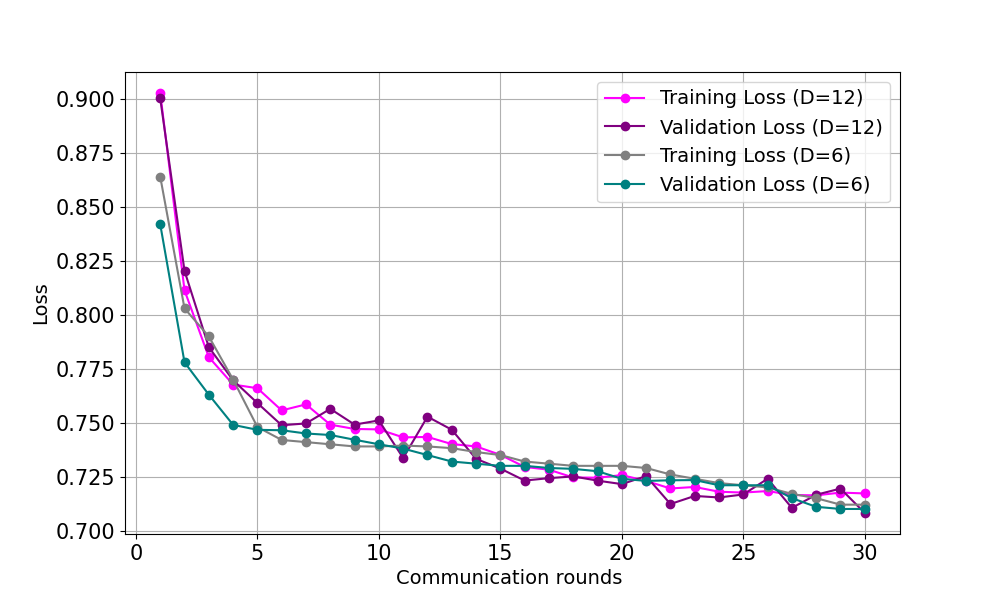}}
\caption{BCE loss convergence over communication rounds}
\label{fig4}\vspace{-4mm}
\end{figure}

\subsection{First stage of FL}
For training a federated model, we use the CAE network architecture described in the previous section. We assume all 12 clients will participate in training the same model locally and further communicating their model weights with the server. The federated server uses the SGD optimizer to aggregate the weights obtained from the local models. Additionally, we intend to train the federated model for a smaller number of communication rounds i.e. 15 to achieve a robust convergence. Fig. \ref{fig3} shows the convergence of Mean Square Loss (MSE) loss function for the number of training samples: 3600 and validation samples: 400, which highlights attaining a faster convergence by stabilizing over 10 communication rounds. This enables the $\mathcal{M}_{ENC+DEC}$ being trained effectively and has captured the temporal representation of the SSB information across all datasets without reflecting any signs of overfitting. This further ensures that $\mathcal{M}_{ENC+DEC}$ is efficiently aggregating the weights from 12 and 6 clients using FedAVG algorithm regardless of non-IID distribution across all datasets, therefore enabling the model's ability to reconstruct the data.

\begin{figure}[t]
\centerline{\includegraphics[scale = 0.28]{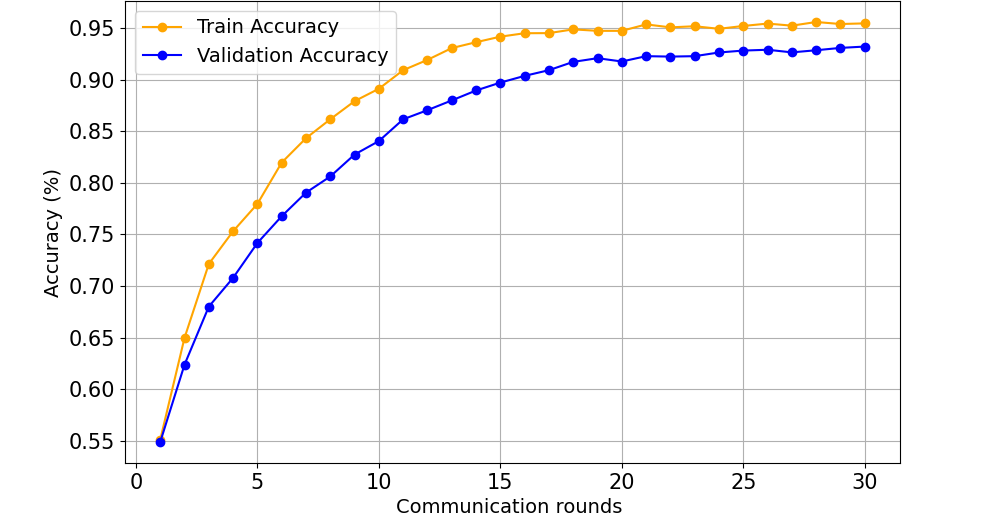}}
\caption{Training and Validation accuracy with 50\% of clients using FedAVG+FedProx}
\label{fig5} \vspace{-4mm}
\end{figure}
\vspace{-3.5mm}
\begin{table}[t]
    \centering
    \caption{Jamming detection using all clients and 50\% of clients with testing accuracy}\label{tab3}
    \resizebox{8cm}{!}{
    \begin{tabular}{|c|c|c|c|c|}
    \hline
        \textbf{Algorithms } & \textbf{Precision} & \textbf{Recall} & \textbf{F1-score} & \textbf{Accuracy} \\ \hline
        FedAVG (D = 12) & 0.65 & 0.79 & 0.71 & 0.62 \\ 
        FedAVG (D = 6) & 0.75 & 0.86 & 0.8 & 0.75 \\ 
        FedProx (D = 12) & 0.72 & 0.86 & 0.8 & 0.75 \\ 
        FedProx (D = 6) & 1 & 0.67 & 0.8 & 0.8 \\ 
        FedAVG + FedProx (D = 12) (Proposed) & 0.97 & 0.91 & 0.90 & 0.89 \\ 
        FedAVG + FedProx (D = 6) (Proposed) & 0.94 & 0.90 & 0.92 & 0.92 \\ \hline
    \end{tabular}
    }\vspace{-0.4cm}\
\end{table}

\subsection{Second stage of FL}
In the second stage of federated learning (FL), the model $\mathcal{M}_{ENC+FCN}$ focuses on binary classification, optimizing the Binary Cross Entropy loss function using the Adam optimizer. We assume 12 clients and 50\% of total clients (i.e 6) are participating during each communication round of FedProx algorithm. Fig. \ref{fig4} depicts the BCE loss convergence of the model trained and validated over 30 communication rounds. This shows that the model generalizes well without significant overfitting. The model exhibits strong performance, as evidenced by high training and validation accuracy, both converging after around 15 communication rounds. The model reaches a stable, well-generalized state, achieving approximately 93\% validation accuracy by the end of the training process as highlighted in Fig. \ref{fig5}. This performance shows that using the FedAVG (first stage) + FedProx (second stage) algorithm with 50\% client participation leads to efficient and accurate federated learning, with consistent results across both training and validation sets.
The performance results presented in Table \ref{tab3} highlight the differences in jamming detection across various algorithms: FedAVG, FedProx, and the proposed two-stage FL approach using all 12 clients (datasets) and a random sampling of 6 clients. When comparing FedAVG across these setups, the detection performance improves when only 50\% of the clients are used, with increases in precision, recall, F1-score, and accuracy. FedAVG with 6 clients achieves better results with accuracy of 0.75 than using all clients with accuracy of 0.62, indicating that a smaller subset of clients performs better in detecting jammed signals. For FedProx, while it outperforms FedAVG when using all 12 clients, achieving higher recall and accuracy (0.75), the performance drops with 50\% client participation, showing reduced recall despite achieving perfect precision. However, the proposed method (FedAVG + FedProx) consistently yields the best results in both setups. With all 12 clients, the proposed approach reaches high precision (0.97), recall (0.91), F1-score (0.90), and accuracy (0.89). Even with 50\% of the clients, the proposed method maintains high metrics, showing robustness and scalability with an accuracy of 0.92. Overall, the proposed approach demonstrates superior performance and resilience in detecting jammed signals, even when fewer clients participate in the federated learning process.

\section{Conclusion}\label{C}
We have investigated the efficient jamming detection in 5G networks by proposing a two-stage federated learning framework that integrates unsupervised learning through a convolutional autoencoder (CAE) in the first stage and a supervised classification model in the second stage. By combining the FedAVG and FedProx algorithms, the framework addresses challenges posed by non-IID data across distributed clients, ensuring data privacy and efficient learning. Experimental results demonstrate that the proposed method achieves superior performance, with high precision, recall, F1-scores, and accuracy, particularly when using 50\% of clients in each communication round. The results confirm that the framework detects jamming attacks effectively and maintains robustness and scalability in real-world 5G environments, providing a promising solution for enhancing cybersecurity in heterogeneous 5G networks.

\section*{Acknowledgment}
This work was supported in part 
by the Natural Sciences and Engineering Research Council of Canada (NSERC) under the DISCOVERY and CREATE TRAVERSAL Programs.

\bibliographystyle{IEEEtran}

\end{document}